# Towards Causal Interpretation of Sexual Orientation in Regression Analysis: Applications and Challenges

**Author list**

Junjie Lu, MD, MS, MPH, Department of Epidemiology and Population Health, Stanford University School of Medicine, Stanford, CA, USA (email: junjielu@stanford.edu)

Zhongyi Guo, BS, Department of Epidemiology and Population Health, Stanford University School of Medicine, Stanford, CA, USA (email: guozy@stanford.edu)

David H. Rehkopf, ScD, MPH, Department of Epidemiology and Population Health, Department of Medicine (Division of Primary Care and Population Health), Department of Pediatrics, Department of Health Policy, Stanford University School of Medicine, Department of Sociology, Stanford University, Stanford, CA, USA (email: drehkopf@stanford.edu)

**Corresponding author**

David H. Rehkopf, ScD, MPH, Department of Epidemiology and Population Health, Department of Medicine (Division of Primary Care and Population Health), Department of Pediatrics, Department of Health Policy, Stanford University School of Medicine, Department of Sociology, Stanford University, Stanford, CA, USA (email: drehkopf@stanford.edu)

**Abstract**

This study presents an approach to analyze health disparities in Sexual and Gender Minority (SGM) populations, with a focus on the role of social support levels as an example to allow causal interpretations of regression models. We advocate for precisely defining the exposure variable and incorporating mediators into analyses, to address the limitations of comparing counterfactual outcomes solely between SGM and heterosexual populations. We define sexual orientation into domains (attraction, behavior, and identity), and emphasize a consideration of these elements either separately or together, depending on the research question. We also introduce social support measured before and after the disclosure of sexual orientation to facilitate inference. We illustrate this approach by examining the association between SGM status and depression diagnosis with data from the 2020 and 2021 National Health Interview Survey. We find a direct effect of SGM status on depression (OR: 3.07, 95% CI: 2.64 - 3.58) and no indirect effect throughsocial support (OR: 1.07, 95% CI: 0.87-1.31). Our research emphasizes the necessity of the comprehensive measurement of sexual orientation and a focus on intervenable variables like social support in order to empower SGM communities and address SGM related health inequalities.

**Acknowledge**

Author affiliations: Department of Epidemiology and Population Health, Stanford University School of Medicine, Stanford, California, United States (Junjie Lu, Zhongyi Guo, David H. Rehkopf ); Department of Medicine (Division of Primary Care and Population Health), Department of Pediatrics, Department of Health Policy, Stanford University School of Medicine, Department of Sociology, Stanford University, Stanford, California, United States (David H. Rehkopf).

**Towards Causal Interpretation of Sexual Orientation in Regression Analysis: Applications and Challenges**

**Abstract**

This study presents an approach to analyze health disparities in Sexual and Gender Minority (SGM) populations, with a focus on the role of social support levels as an example to allow causal interpretations of regression models. We advocate for precisely defining the exposure variable and incorporating mediators into analyses, to address the limitations of comparing counterfactual outcomes solely between SGM and heterosexual populations. We define sexual orientation into domains (attraction, behavior, and identity), and emphasize a consideration of these elements either separately or together, depending on the research question. We also introduce social support measured before and after the disclosure of sexual orientation to facilitate inference. We illustrates this approach by examining the association between SGM status and depression diagnosis with data from the 2020 and 2021 National Health Interview Survey. We find a direct effect of SGM status on depression (OR: 3.07, 95% CI: 2.64 - 3.58) and no indirect effect through social support (OR: 1.07, 95% CI: 0.87-1.31). Our research emphasizes the necessity of the comprehensive measurement of sexual orientation and a focus on intervenable variables like social support in order to empower SGM communities and address SGM related health inequalities.

In epidemiological research, the counterfactual outcome approach is an important methodology for understanding health disparities. This approach involves imagining two scenarios: one where a hypothetical intervention is applied to a population, and the other where it is not. By comparing the health outcomes in these two scenarios, researchers can assess the potential impact of the intervention (1).

Several approaches have been pursued to integrate counterfactual methods into social epidemiology, including redefining exposure variables, differentiating exposure from the hypothetical intervention, and considering multiple potential interventions (2). For instance, Tyler J. VanderWeele and colleagues have proposed an approach to assess the effect of race by being specific about the exposure and incorporating socioeconomic status at different life stages in regression models, treating these factors as hypothetical interventions for causal analysis (3).

The use of a counterfactual framework in causal inference to investigate health disparities between SGM and heterosexual groups has been less developed. Commonly, researchers include sexual orientation in their regression models (4–9), and under a traditional counterfactual framework, the coefficient representing SGM status would reflect health disparities between an exclusively SGM population and a heterosexual one. This suggests a hypothetical intervention capable of altering individuals' SGM status, which is unfeasible. Therefore, we need to formally extend the causal interpretation practice to address the gap related to health disparities between SGM and heterosexual populations.

Multiple intervenable strategies are available to enhance the health of the SGM community, including education program, community-based support, and comprehensive health initiatives (10–12). We choose social support as an example because it is an integral component derived from one's social connections, and manifests in various interconnected forms (13–15). Social support is particularly important for SGM populations who often face stigma and marginalization, leading to significant health challenges (16–18). Discrimination at both societal and policy levels can worsen health outcomes, leading to emotional distress, substance misuse, and higher suicide rates (16,19,20). By incorporating social support into causal inference models related to SGM health, we can evaluate the effects of societal and policy shifts on the health of SGM populations.

In this illustrative case, we tested the association between SGM status and depression, focusing on the role of social support levels. Several pieces of evidence underscore the vulnerability of SGM populations to depression (16,21,22). The factors contributing to the elevated risk of depression include the chronic stress from family rejection, societal stigmatization, and minority stress (23–26). However, the potential effect of social support levels as actionable variables from a causal standpoint is understudied. That is, what will the disparity in depression be, if we set the social support level same between SGM and heterosexual populations?

To address these limitations in the social epidemiologic literature, our study aims to: a) propose a causal inference approach to explore health inequalities in the SGM population, using social support as an example of an intervenable variable; b) conduct an illustrative case study by examining the association between SGM status and depression using a nationally representative dataset, focusing on how disparities in depression might be altered if the proposed causal approach were applied to social support levels.

## METHODS

**Domains of sexual orientation**

Sexual orientation encompasses three key domains: sexual attraction, behavior, and identity (27–29). Attraction encompasses emotional, romantic, or sexual feelings towards other people and exists on a spectrum that includes heterosexual, homosexual, bisexual, and other categories. Behavior refers to the physical acts, which may not always correspond with one's attractions or identity, often influenced by societal pressures. Identity is how individuals label their sexual orientation, shaped by personal, societal, and cultural experiences, with common labels like gay, lesbian, bisexual, and straight, which may not always align with attraction and behavior. Most population-based studies have not measured all three, yet their separate or collective analysis could affect causal interpretations. A brief overview of these dimensions is documented in Appendix S1.

**Social support levels for SGM as intervenable variables**

We hypothesized social support as an actionable target for interventions, in contrast with sexual orientation. We categorized social support into two types: social support $t_0$— the support received before disclosing sexual orientation, including factors like legal acceptance of same-sex marriage and family support, and social support $t_1$, which was the support after the disclosure. Social support $t_0$ and social support $t_1$ varied; for example, SGM individuals in LGBTQ-friendly areas might receive strong support at both time, but in other contexts, social support $t_0$ might be low initially and increase after legalization of same-sex marriage.

The distinction between the social support $t_0$ and the social support $t_1$ was vital for our counterfactual causal analysis of the health outcome (depression in the illustrative study). The former was related to non-mediated variables, while the latter was a mediated variable between the sexual orientation and the health outcomes. Our analysis of sexual orientation's impact on health considered both direct effects and the mediating role of the social support $t_1$.

**Causal interpretation: only control for social support $t_0$**

We used Directed Acyclic Graph (DAG) to illustrate the hypothetical scenarios. The primary aim of employing DAGs was to define the exposure variable precisely by showing the backdoor paths between the exposure and the outcome to clarify the causal relationship of interest (30). To enhance interpretability, we consolidated the diverse subgroups within the SGM population into a unified category. This approach facilitated an easier comparative analysis with the heterosexual population and could be adapted for more detailed examinations of SGM subgroups as needed. We also wanted to note that interactions existed between the domains of sexual orientation. We depicted sexual identity as the common effect of sexual attraction and behavior from a developmental perspective where individuals might define their identity through these experiences (31,32).

*Sexual orientation as different domains*

One approach separated sexual orientation into three distinct domains. This approach was beneficial because it allowed us to examine how each specific domain was associated with the outcome. For example, if we wanted to examine the causal effect of sexual behavior on health outcomes, not accounting for other sexual orientation domains and social support $t_0$ could introduce bias. Specifically, as shown in Figure 2 Panel A: the environmental, social, and genetic factors (H) were linked to three domains and the social support $t_0$, and all of which affected health outcomes (Y) (33). If we did not control the sexual attraction, there would be a backdoor path between the sexual behavior and the outcome: from "sexual behavior" to "H" to "sexual attraction" and to "Y" (30). Similarly, neglecting to account for social support $t_0$ could cause confounding as well: from "sexual behavior" to "H" to "social support $t_0$" and to "Y". Therefore, to examine the relationship between sexual behavior and health outcomes, controlling sexual attraction and social support $t_0$ was required.

*Sexual orientation as joint effect*

The other approach was to evaluate the combined effect of sexual orientation as a unified primary exposure. This approach could be useful when not all domains were measured. To minimize confounding, the model should also incorporate social support $t_0$. Figure 2 Panel B illustrated this relationship.

####################
#Figure 1 Panel A & B#
####################

*Interpretation using social support $t_0$*

Instead of focusing on hypothetical counterfactual setting of sexual orientation, we focus instead on social support $t_0$. The assumptions for this approach were: (a) Consistency: The coefficient for the social support $t_0$ consistently reflected its impact on the outcome. This assumption could be violated if the effect of social support $t_0$ were different between the counterfactual and the observed scenario; (b) Exchangeability: Once adjusted for the social support $t_0$ and other confounding factors, there would be no indirect paths that existed from the exposure to health outcomes. These assumptions could be violated if there were unmeasured and uncontrolled confounding. Noted that these assumptions were essentially the same as the no mediator-outcome confounding in causal mediation analysis (34).

With these assumptions, the coefficient for sexual orientation from the regression model could be interpreted as the health inequality that would remain between SGM and heterosexual populations, if the SGM population's social support $t_0$ were set equal to that of the heterosexual populations. In contrast, in the approach we propose here, the inference was not on sexual orientation *per se* but on an intervenable variable, instead, social support $t_0$, which offered more actionable insights for programs and policy. Detailed proof is provided in Appendix S2.

**Causal interpretation: control for mediated variables social support $t_1$**

In assessing the impact of sexual orientation on health outcomes, the inclusion of social support $t_1$ in the regression model was dependent on the analysis goal. If we aim to understand the total effect of sexual orientation, social support $t_1$ should not be included in order to prevent blocking the causal pathway to the health outcome. Conversely, to evaluate how social support $t_0$ directly explained health inequalities, it should be included. Figure 2 Panel A demonstrates this concept, distinguishing between the mediated effects (blue lines) of sexual orientation elements through the social support $t_1$ and the direct effects (red lines) not through the social support $t_1$. Figure 2 Panel B simplified this by considering the joint effect of sexual orientation elements as the exposure.

For the mediated interpretation, similar assumptions were needed: (a) Consistency: the social support $t_1$'s coefficient should indicate its effect on the health outcome; (b) Exchangeability: After accounting for the social support $t_0$ and other confounding factors, the social support $t_1$ should have no indirect pathways to the health outcome, ensuring there was no confounding between mediators and outcomes.

########################
##Figure 2 Panel A & B##
########################

*Direct effect*

With these assumptions, the direct effect of sexual orientation on the health outcome not through social support $t_1$ could be interpreted as follows: the health inequalities that remain for a population with a particular social support $t_0$, if within this population, social support $t_1$ of the SGM members were set equal to that of the heterosexual members. The proof is shown in Appendix S2.

*Indirect effect*

Similarly, the indirect effect of sexual orientation on the health outcome acting through social support $t_1$ could be interpreted as follows: the difference in the health outcome for a SGM population with a particular social support $t_0$, if their social support $t_1$ were set equal to heterosexual members' social support $t_1$. The proof is shown in Appendix S2.

*Total effect*

The total effect for those with a particular social support $t_0$ was equal to the sum of the direct and indirect effect. The proof is shown in Appendix S2.

**Illustrative case study**

We offered an illustrative example that demonstrated how to interpret the coefficient of SGM status as a joint effect in regression models by focusing on the role of social support, using data from the National Health Interview Survey (NHIS).

*Study design*

The National Health Interview Survey (NHIS), conducted by the National Center for Health Statistics (NCHS), is a cross-sectional household interview survey that provides nationally representative data on the health of the U.S. civilian, noninstitutionalized population. Its primary purpose is to track health trends by gathering and analyzing information on various health topics (35).

*Population and sample characteristics*

NHIS includes the civilian, noninstitutionalized population of the United States, in all 50 states and the District of Columbia. The survey employed geographically clustered sampling to select residential units for participation (35). Our analysis included data from the 2020 and 2021 NHIS waves of data collection to teste the association between sexual orientation and depression diagnosis. We focused on 2020 and 2021 because these years included key variables related to social support (36).

*Exposure*

Sexual orientation was determined based on responses to the NHIS question: "Do you think of yourself as gay/lesbian; straight, that is, not gay/lesbian; bisexual; something else; or you don't know the answer?" There was a non-response option "Refused." Respondents were divided into two groups: the SGM group, consisting of individuals who identified as gay/lesbian, bisexual, or another non-straight orientation, and the straight group, consisting of individuals who identified as straight and served as the reference group.

*Outcome*

The outcome, depression diagnosis, was defined using the question, "Have you ever been told by a doctor or other health professional that you had any type of depression?" with the options "Yes" or "No".

*Covariates*

For the illustrative purpose, the family poverty ratio was included in the analysis as a proxy for the social support $t_0$, where a ratio of 1 indicates income at the poverty level, and a ratio greater than 1 indicates income above the poverty level. We use family poverty ratio as a proxy of social support $t_0$ because people in poverty may face social isolation and reduced access to supportive networks, which can exacerbate the challenges associated with low income (37,38). The social support $t_1$ was measured by two questions. The first assessed the frequency of received support: "How often do you get the social and emotional support you need?" with options including "Always", "Usually", "Sometimes", "Rarely" and "Never." The second question assessed changes in support levels over the past year: "Compared with 12 months ago, would you say that you now receive more social and emotional support, less social and emotional support, or about the same?" In both instances, the most favorable responses ("Always" and "More social and emotional support") were defined as reference categories.

Other covariates included age, biological sex assigned at birth, race and education level. Age was measured as a continuous variable, and the biological sex was categorized as either male or female, with male as the reference category. Race was classified into categories of White, Black or African American, Asian, American Indian or Alaska Native (AIAN), and Other. We acknowledge the limitations of an "other" category and the exclusion of Multiracial individuals from analyses, as the aforementioned categories represented the most detailed level of data available to us. The choice of reference category followed guidelines for reporting of race and ethnicity in medical and science journals (39). Specifically, we coded the variable for race as a categorical variable and avoided collective reference to racial minority groups as "non-White." We listed racial categories in alphabetical order instead of ordering by majority. Educational attainment was categorized into three levels, including high school graduate or below, bachelor's degree, and master's degree and above. The lowest level of educational attainment was used as the reference category.

*Statistical analysis*

We applied multiple imputation using the predictive mean matching method, generating five imputed datasets to address missing data. We used outcome logistic regression models to conduct the effect estimation (40). Specifically, the total effect of SGM status on depression was estimated from a model excluding the social support $t_1$. The direct effect was examined by including the social support $t_1$ in the model. The indirect effect was derived by calculating the difference between the total and direct effect estimates.

To evaluate the influence of unmeasured confounding, we calculated the E-value, which quantified the minimum strength of association that an unmeasured confounder would need to have with both the exposure and outcome to fully explain away a specific treatment-outcome association (41). Additional robustness checks included: 1) modeling without interaction terms; 2) utilizing propensity scores to account for all covariates; 3) applying inverse probability weighting to adjust for confounding. Analyses were weighted according to NHIS guidelines to ensure national representativeness. The report of the analysis followed the STROBE guidelines on cross-sectional studies (Table S3) (42). All tests were two-sided, and the 95% confidence interval excluding 1 was considered statistically significant. Analyses were performed using R (Version 4.1.2). The data and code presented in this study are openly available in GitHub: https://github.com/andersonjunjielu/Causal-interpretation-for-SGM-research

**RESULTS**

**Participants and descriptive analysis**

In the 2020 National Health Interview Survey (NHIS), a total of 31,568 individuals participated, followed by 29,482 participants in the 2021 NHIS. The combined data from these two years yielded an initial sample size of 61,050. Within this sample, 15606 participants in 2020 and 3113 in 2021 had missing data for essential study variables or being non-responders, defined as those who provided responses such as "I didn't know" or "I was uncertain." Detailed comparisons of participants with and without these missing values or being respondents or non-responders were provided in Supplementary Material Table S1 and Table S2. After excluding missing values and non-responsive data, the refined dataset comprised a final sample size of 42,331. The analytic sample flow chart is presented in Figure 3.

Table 1 presents the demographic and clinical characteristics of the participants. The data revealed a higher prevalence of depression in the SGM group (42.9643.0%) compared to the heterosexual group (17.3%). SGM individuals reported less consistent social support, with only 39.4% stating they "always" receive it, versus 54.4% in the heterosexual group. Additionally, SGM individuals experienced more variability in social support changes, with higher percentages reporting both increased and decreased levels of support. A descriptive comparison of the propensity scores between SGM participants and their heterosexual counterparts was included in Supplementary Figure S1.

##########
#Figure 3#
##########

#########
#Table 1#
#########

**Main results**

The direct effect was presented as an odds ratio of 3.1 (95% CI: 2.6, 3.6). This indicated that, in a population with both SGM and heterosexual individuals, if the social support $t_1$ of SGM individuals was hypothetically aligned with that of heterosexual individuals, SGM individuals would still have, on average, 3.1 times odds of depression compared to their heterosexual counterparts, holding the family poverty level constant.

The indirect effect was 1.1 (95% CI: 0.9, 1.3). This indicated that, for a SGM population with a particular family poverty, if their social support $t_1$ were set equal to heterosexual members' social support $t_1$, their odds of depression was 1.1 times higher compared to their heterosexual counterparts with the same family poverty level. The non-significance of the coefficient of SGM status indicated that the effect of SGM status on the depression status was not mediated by the social support $t_1$.

The total effect was 3.3 (95% CI: 2.8, 3.8). This implied that, considering both the direct effect of SGM status and the mediated effect through the social support $t_1$, SGM individuals had 3.3 times odds of depression compared to heterosexual individuals at specific family poverty level. The information was shown in Table 2 and Figure 3.

#########
#Table 2#
#########

##########
#Figure 3#
##########

**Sensitivity analysis**

Statistically, we constructed alternative models, including basic outcome regression, regression utilizing propensity scores, and analysis through inverse probability weighting. Additionally, we performed a complete case analysis, where instances of missing data and non-responses were excluded. As indicated in Table 2 and Figure 3, these alternative methodologies consistently produced results akin to our primary analysis (Range of the total, direct and indirect effect: OR: 3.2-4.2; 2.9-3.7; 1.1-1.2).

To evaluate the influence of unmeasured confounding factors on our findings, we calculated E-values. E-values for the total and direct effects were determined to be 3.0 and 2.9, respectively. Such values implied that any unmeasured confounder responsible for explaining away our findings would need to exhibit at least a 3.0 and 2.9 odds ratio association with both the exposure and the outcome, surpassing the influence of the covariates already observed. This suggested that although the presence of unmeasured confounding could not be ruled out, it would require a strong association to negate the observed association entirely. Conversely, the E-value for the indirect effect was calculated as 1.2. This value indicated a high possibility of the observed association to be biased by unmeasured confounding.

**DISCUSSION**

Our study examined the causal interpretation of the coefficient of SGM status within a regression model, building upon Tyler J. VanderWeele and colleagues' foundational work on race-related causal dynamics in regression analysis (3). Our theoretical exploration integrated three domains of sexual orientation to formulate research hypotheses. We advocated for a precise definition of the exposure variable in studies addressing health disparities in SGM populations with the consistency and exchangeability assumptions. Additionally, we emphasized the need to include more intervenable variables such as social support in SGM health research. Incorporating these variables enhances the interpretability of SGM status coefficients, shifting focus to a more practical aspect: the level of social support, a factor that can be influenced by programmatic and policy changes. We also examined both the social support $t_0$ and $t_1$ -- the former preceding and the latter following the assessment of sexual orientation -- to facilitate a causal analysis of SGM status's direct, indirect, and total impacts on health outcomes.

In our illustration, we utilized 2020 and 2021 NHIS data to examine the combined effect of SGM status on depression diagnoses, considering social support $t_1$ as a mediating factor. Our findings indicated a significant direct influence of SGM status on depression, while the indirect effect was not significant. These outcomes remained similar in sensitivity analyses.

Our study established a framework for conducting health research in the SGM community, applying causal inference methods to assess the potential impact of policies that enhance SGM empowerment through social support. Social support is multifaceted, encompassing elements such as family, friends, significant others, emotional, informational and tangible supports (43,44). Consistent access to social support is linked to better mental health outcomes, including lower levels of anxiety, depression, and suicide rates (45,46).

Our research addressed a gap using a counterfactual approach to examine the potential effect of intervenable methods to narrow the health inequality gap between SGM and heterosexual populations. The intervenable approach extends well beyond social support, encompassing policies, such as the legalization of same-sex marriage. Our findings equipped policymakers with analytical tools to predict the effects of these interventions on SGM communities. Currently, fewer than 40 countries have legalized same-sex marriage, contributing to deficits in social support and widening health disparities (47–49). By using the proposed causal interpretation, policymakers can examine how the health inequality would remain if policies that impact social support, including the legalization of the same-sex marriage, were applied to SGM populations.

In the illustrative case, the significant direct effect of SGM status on depression was similar to prior research (21,24,50). However, the mediated effect of social support was not significant. This was possibly because the social support measurement was not comprehensive enough or measurement error was large. Our measurement of social support $t_1$ was limited to only two questions. Other more validated tools, like the Everyday Discrimination Scale (51), were not available in the NHIS dataset. Future studies should consider more reliable and valid measurements for social support.

Our study had several limitations. Firstly, while family poverty might not serve as an ideal proxy for social support prior to the measurement of SGM status, but it was the closest variable available in the dataset. Future research should aim to directly measure social support. Secondly, the illustrative case study relied on self-reported depression diagnoses, which could lead to outcome misclassification. Future research could employ more objective depression assessments. Third, challenges were encountered in measuring sexual orientation, particularly in national studies like NHANES, which does not often measure all domains of sexual orientations (52). Besides, there could be interactions between the domains of sexual orientation. We assumed that the sexual identity was the common effect of sexual attraction and behavior from a developmental perspective and for easier mathematics proof. Researchers should be specific about the assumption of the relationship between domains. Furthermore, although some large-scale surveys measure all sexual orientation domains, they are exclusively within LGBTQ communities, thereby complicating comparative health assessments between LGBTQ and heterosexual populations (53). Thus, we recommend nationally representative studies that encompass all domains of sexual orientation. Another measurement challenge was the dynamic nature of sexual orientation. Considering its potential evolution and fluidity over a person's lifetime, a single assessment of sexual orientation may be insufficient (54). Consequently, longitudinal surveys that repeatedly record participants' sexual orientation would be informative.

In conclusion, we introduced a counterfactual framework for interpreting the coefficient of SGM status within regression models, particularly through intervenable variables, using social support as an example. This approach emphasized accurately defining sexual orientation and associated domains as exposure variables. Through the illustrative case study, we offered policymakers analytical tools to assess how social policies could enhance social support and reduce health disparities between SGM and heterosexual populations.

**Table 1.** Demographic and Clinical Characteristics of Participants in the National Health Interview Survey (NHIS), 2020-2021, Stratified by Sexual and Gender Minority (SGM) Status

|  | SGM | Heterosexual |
|---|---|---|
| **Age, mean (SD)** | 42.62 (17.30) | 54.06 (18.17) |
| **Sex, n (%)** | | |
| Male | 801 (42.58) | 18491 (45.71) |
| Female | 1080 (57.42) | 21959 (54.29) |
| **Race, n (%)** | | |
| American Indian or Alaska Native | 59 (3.14) | 676 (1.67) |
| Asian | 74 (3.93) | 2455 (6.07) |
| Black or African American | 188 (9.99) | 4480 (11.08) |
| Other | 67 (3.56) | 545 (1.35) |
| White | 1493 (79.37) | 32294 (79.84) |
| **Educational Attainment, n (%)** | | |
| High School Graduate or Below | 389 (20.68) | 12542 (31.00) |
| Bachelor's Degree | 1125 (59.81) | 21410 (52.93) |
| Master's degree and above | 367 (19.51) | 6498 (16.06) |
| **Family poverty ratio, mean (SD)** | 4.29 (3.09) | 4.39 (3.00) |
| **Frequency of Social Support, n (%)** | | |
| Always | 741 (39.39) | 22008 (54.41) |
| Usually | 649 (34.50) | 10583 (26.16) |
| Sometimes | 323 (17.17) | 4680 (11.57) |
| Rarely | 109 (5.79) | 1469 (3.63) |
| Never | 59 (3.14) | 1710 (4.23) |
| **Changes in social support levels, n (%)** | | |
| More social support | 387 (20.57) | 4905 (12.13) |
| Less social support | 355 (18.87) | 4537 (11.22) |
| About the same | 1139 (60.55) | 31008 (76.66) |
| **Depression, n (%)** | | |
| Yes | 808 (42.96) | 6997 (17.30) |
| No | 1073 (57.04) | 33453 (82.70) |

**Table 2.** Total, Direct, and Indirect Effects of the Association Between Sexual and Gender Minority (SGM) Status and Depression Status in the National Health Interview Survey (NHIS), 2020-2021 [a]

|  | **Primary analysis** [e] | **Simple outcome regression** [f] | **Propensity score regression** | **Inverse possibility weighting** |
|---|---|---|---|---|
| **Multiple Imputation** | | | | |
| **Total effect** [b] | 3.3 (2.8, 3.8) | 3.7 (3.1, 4.4) | 3.7 (3.3, 4.2) | 4.1 (3.6, 4.6) |
| **Direct effect** [c] | 3.1 (2.6, 3.6) | 3.4 (2.8, 4.0) | 3.3 (3.0, 3.7) | 3.7 (3.3, 4.1) |
| **Indirect effect** [d] | 1.1 (0.9, 1.3) | 1.1 (0.9, 1.4) | 1.1 (1.0, 1.3) | 1.1 (1.0, 1.3) |
| **Complete Data Analysis** | | | | |
| **Total effect** | 3.2 (2.8, 3.8) | 3.8 (3.4, 4.3) | 3.8 (3.4, 4.3) | 4.2 (3.7, 4.7) |
| **Direct effect** | 2.9 (2.5, 3.5) | 3.3 (2.9, 3.8) | 3.2 (2.9, 3.7) | 3.5 (3.1, 4.0) |
| **Indirect effect** | 1.1 (0.9, 1.4) | 1.2 (1.0, 1.4) | 1.2 (1.0, 1.4) | 1.2 (1.0, 1.4) |

a. All models adjusted for family poverty, age, biological sex, race, ethnicity, educational attainment, and survey year.
b. Total Effect: The impact of sexual and gender minority (SGM) status on depression status, without adjusting for social support t1(measured after the disclosure of SGM status).
c. Direct Effect: The influence of SGM status on depression prevalence, accounting for subsequent social support.
d. Indirect Effect: Calculated as the difference between the total and direct effects, this reflects the mediated effect by social support t1 (measured after the disclosure of SGM status).
e. Primary analysis: The models centered all covariates and examined the relationship between depression status and SGM status, including interaction terms between SGM status and the centered covariates.
f. Simple regression analysis: The models did not include the interaction terms between SGM status and the covariates.

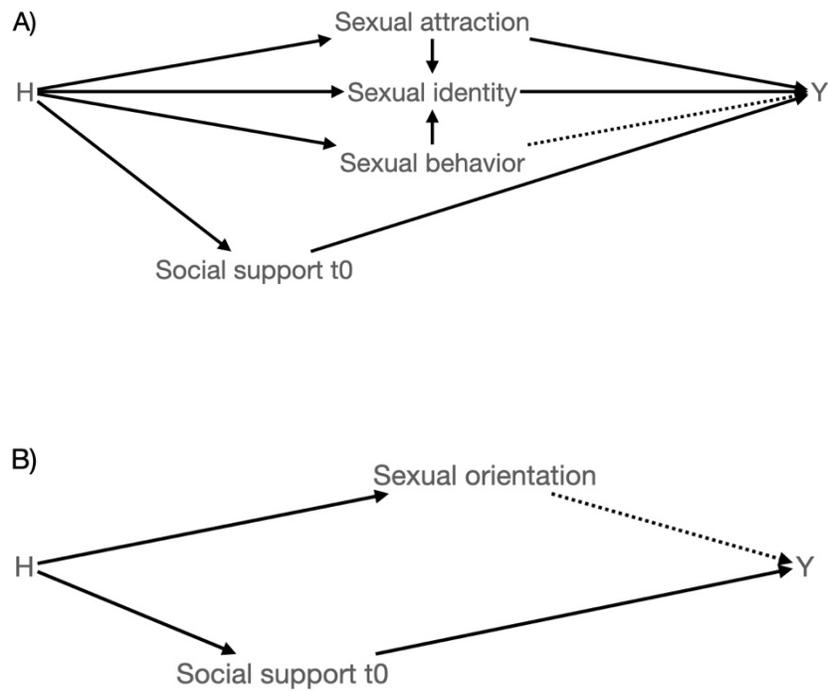

**Figure 1.** Directed Acyclic Graph illustrating the relationship between sexual orientation and health outcomes: A) Sexual orientation as distinct elements. "H" denoted environmental, social, and genetic factors, while "Y" denoted the health outcomes under study. Dashed line denoted the causal relationship of interests. When examining the causal effect of sexual behavior on health outcomes, not accounting for sexual attraction and social support t0 (measured before the disclosure of SGM status) can introduce bias; B) Sexual orientation in terms of its combined effect. "H" denotes environmental, social, and genetic factors, while "Y" denoted the health outcomes under study. When examining the causal effect of sexual orientation as a joint effect on health outcomes, not accounting for baseline social support (measured before the disclosure of SGM status) could introduce bias.

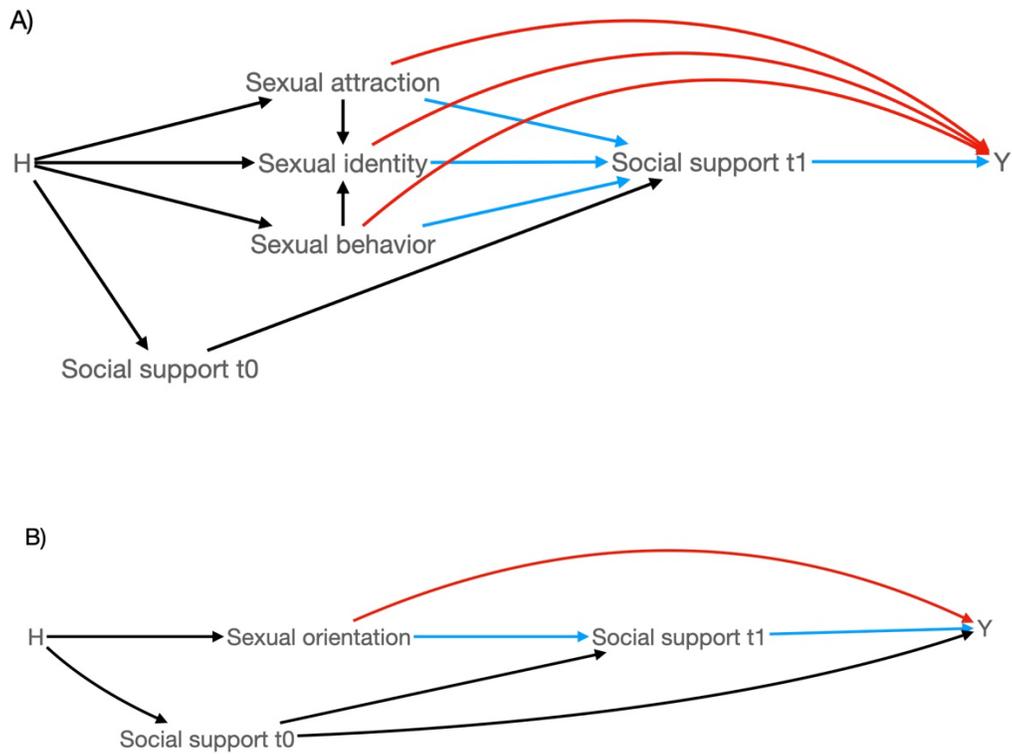

**Figure 2.** Directed Acyclic Graph depicting the association between sexual Orientation, social supports, and health outcomes: A) Sexual orientation as distinct elements. "H" denoted environmental, social, and genetic factors, while "Y" denoted the health outcomes under study. Red lines denoted as the direct effect of sexual orientation elements on the health outcome. Blue lines denoted as the indirect effect of sexual orientation elements on the health outcome mediated by the social support t1 (measured after the disclosure of SGM status). B) Sexual orientation in terms of its combined effect. "H" denoted environmental, social, and genetic factors, while "Y" denoted as the health outcomes under study. Red lines denoted the direct effect of sexual orientation as a joint effect on the health outcome. Blue lines denoted as the indirect effect of sexual orientation as a joint effect on the health outcome mediated by the social support t1 (measured after the disclosure of SGM status).

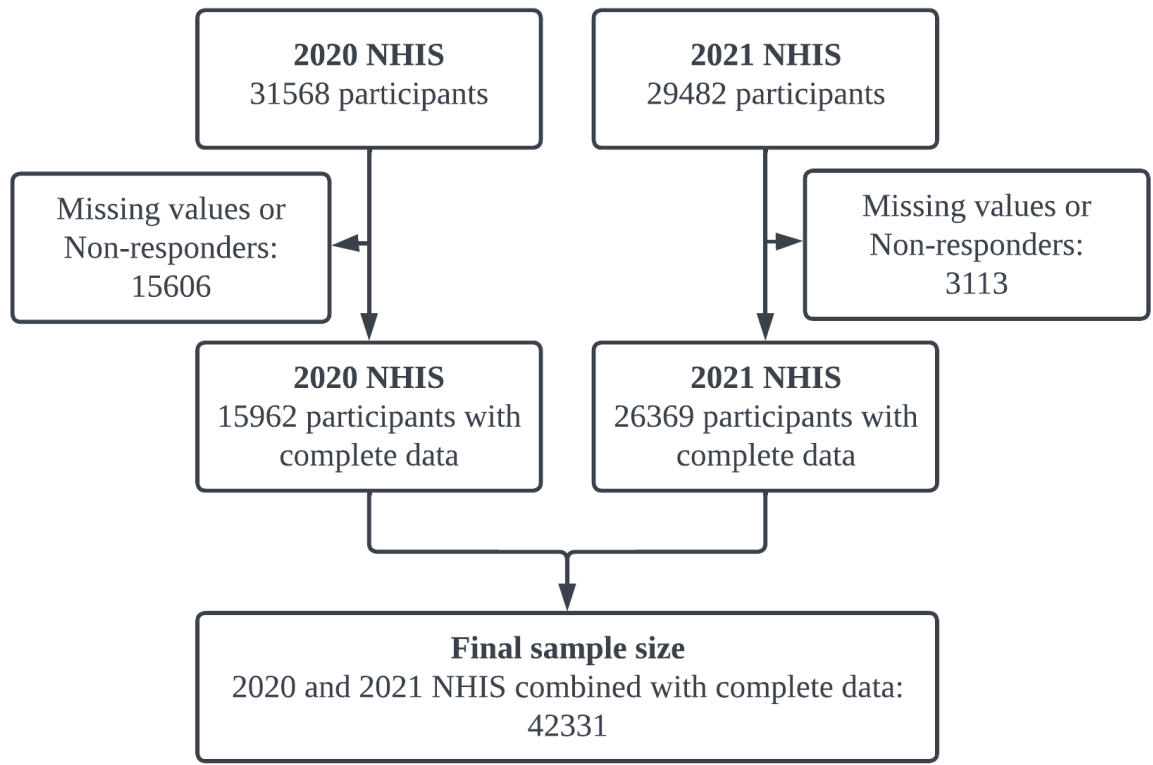

**Figure 3.** Analytic sample flowchart for the 2020 and 2021 National Health Interview Survey (NHIS)

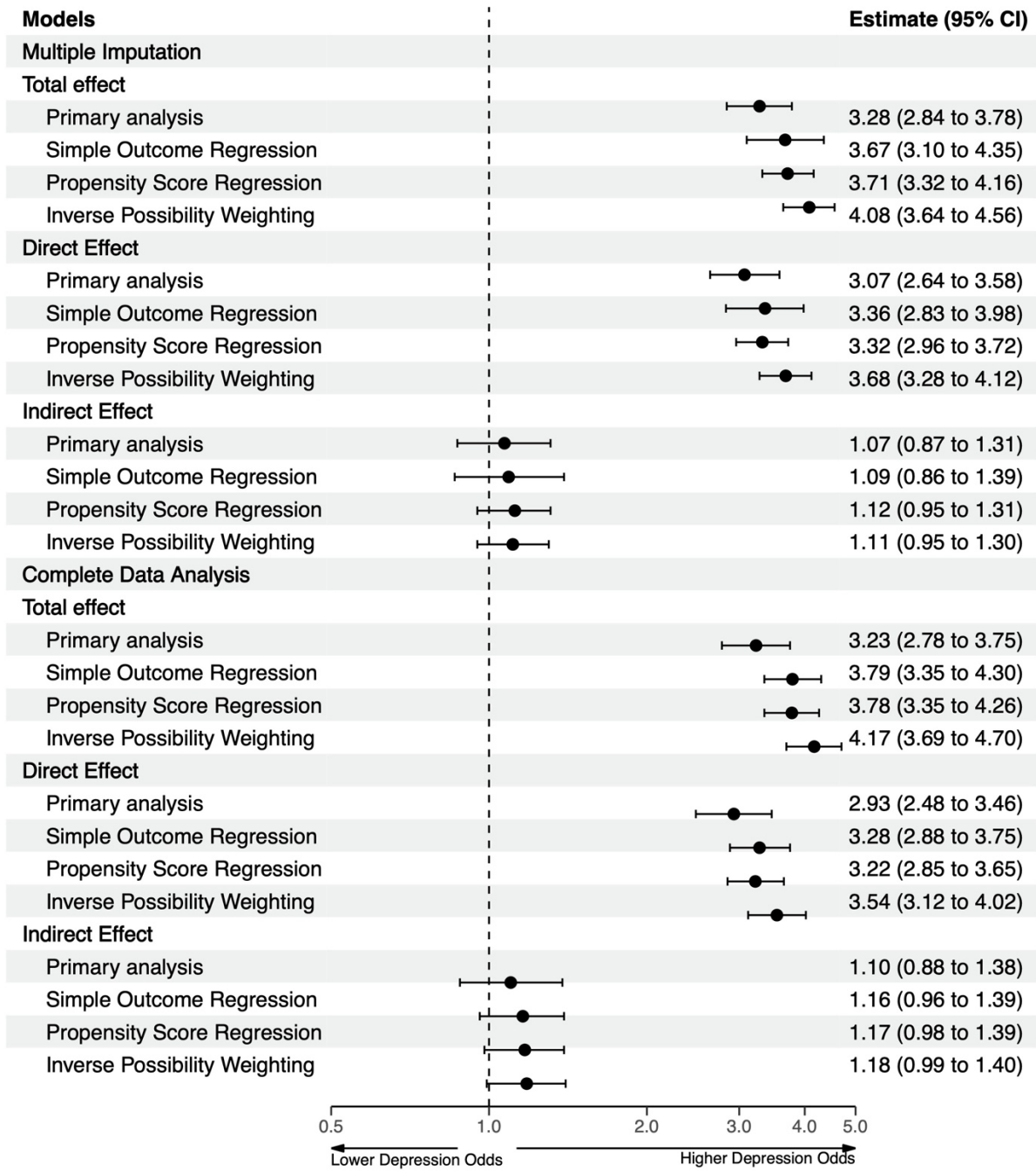

**Figure 3.** Forest plot of the direct and indirect effects of sexual and gender minority (SGM) status on depression, mediated by social support level: All models were adjusted for family poverty, age, biological sex, race/ethnicity, educational attainment, and survey year. Total effect represented the impact of SGM status on depression status, without adjusting social support t1 (measured after the disclosure of SGM status). Direct effect denoted the influence of SGM status on depression prevalence, accounting for social support t1 (measured after the disclosure of SGM status). Indirect effect was calculated as the difference between the total and direct effects, this reflected the mediated effect by the social support t1 (measured after the disclosure of SGM status). Primary analysis centered all covariates and examined the relationship between depression status and SGM status, including interaction terms between SGM status and the centered covariates. Simple regression analysis models did not include the interaction terms between the SGM status and the covariates.

**Table 1.** Demographic and Clinical Characteristics of Participants in the National Health Interview Survey (NHIS), 2020-2021, Stratified by Sexual and Gender Minority (SGM) Status

|  | SGM | Heterosexual |
|---|---|---|
| **Age, mean (SD)** | 42.62 (17.30) | 54.06 (18.17) |
| **Sex, n (%)** | | |
| Male | 801 (42.58) | 18491 (45.71) |
| Female | 1080 (57.42) | 21959 (54.29) |
| **Race, n (%)** | | |
| American Indian or Alaska Native | 59 (3.14) | 676 (1.67) |
| Asian | 74 (3.93) | 2455 (6.07) |
| Black or African American | 188 (9.99) | 4480 (11.08) |
| Other | 67 (3.56) | 545 (1.35) |
| White | 1493 (79.37) | 32294 (79.84) |
| **Educational Attainment, n (%)** | | |
| High School Graduate or Below | 389 (20.68) | 12542 (31.00) |
| Bachelor's Degree | 1125 (59.81) | 21410 (52.93) |
| Master's degree and above | 367 (19.51) | 6498 (16.06) |
| **Family poverty ratio, mean (SD)** | 4.29 (3.09) | 4.39 (3.00) |
| **Frequency of Social Support, n (%)** | | |
| Always | 741 (39.39) | 22008 (54.41) |
| Usually | 649 (34.50) | 10583 (26.16) |
| Sometimes | 323 (17.17) | 4680 (11.57) |
| Rarely | 109 (5.79) | 1469 (3.63) |
| Never | 59 (3.14) | 1710 (4.23) |
| **Changes in social support levels, n (%)** | | |
| More social support | 387 (20.57) | 4905 (12.13) |
| Less social support | 355 (18.87) | 4537 (11.22) |
| About the same | 1139 (60.55) | 31008 (76.66) |
| **Depression, n (%)** | | |
| Yes | 808 (42.96) | 6997 (17.30) |
| No | 1073 (57.04) | 33453 (82.70) |

**Table 2.** Total, Direct, and Indirect Effects of the Association Between Sexual and Gender Minority (SGM) Status and Depression Status in the National Health Interview Survey (NHIS), 2020-2021 [a]

|  | **Primary analysis** [e] | **Simple outcome regression** [f] | **Propensity score regression** | **Inverse possibility weighting** |
|---|---|---|---|---|
| **Multiple Imputation** | | | | |
| **Total effect** [b] | 3.3 (2.8, 3.8) | 3.7 (3.1, 4.4) | 3.7 (3.3, 4.2) | 4.1 (3.6, 4.6) |
| **Direct effect** [c] | 3.1 (2.6, 3.6) | 3.4 (2.8, 4.0) | 3.3 (3.0, 3.7) | 3.7 (3.3, 4.1) |
| **Indirect effect** [d] | 1.1 (0.9, 1.3) | 1.1 (0.9, 1.4) | 1.1 (1.0, 1.3) | 1.1 (1.0, 1.3) |
| **Complete Data Analysis** | | | | |
| **Total effect** | 3.2 (2.8, 3.8) | 3.8 (3.4, 4.3) | 3.8 (3.4, 4.3) | 4.2 (3.7, 4.7) |
| **Direct effect** | 2.9 (2.5, 3.5) | 3.3 (2.9, 3.8) | 3.2 (2.9, 3.7) | 3.5 (3.1, 4.0) |
| **Indirect effect** | 1.1 (0.9, 1.4) | 1.2 (1.0, 1.4) | 1.2 (1.0, 1.4) | 1.2 (1.0, 1.4) |

a. All models adjusted for family poverty, age, biological sex, race, ethnicity, educational attainment, and survey year.
b. Total Effect: The impact of sexual and gender minority (SGM) status on depression status, without adjusting for social support t1(measured after the disclosure of SGM status).
c. Direct Effect: The influence of SGM status on depression prevalence, accounting for subsequent social support.
d. Indirect Effect: Calculated as the difference between the total and direct effects, this reflects the mediated effect by social support t1 (measured after the disclosure of SGM status).
e. Primary analysis: The models centered all covariates and examined the relationship between depression status and SGM status, including interaction terms between SGM status and the centered covariates.
f. Simple regression analysis: The models did not include the interaction terms between SGM status and the covariates.

**Supplementary Material**

**Appendix S1.** A brief introduction about sexual orientation and three key domains: sexual attraction, behavior, and identity

**Appendix S2.** Proof

**Table S1.** Comparison of Demographic and Clinical Characteristics Between Participants with and without Missing Values in the National Health Interview Survey (NHIS), 2020-2021

**Table S2.** Comparison of Demographic and Clinical Characteristics Between Respondents and Non-responses in the National Health Interview Survey (NHIS), 2020-2021

**Table S3.** STROBE Check List

**Figure S1.** Comparison of Propensity Scores Between Sexual and Gender Minority (SGM) and Heterosexual (Straight) Participants in the National Health Interview Survey (NHIS), 2020-2021, With and Without the Inclusion of Subsequent Social Support in the Propensity Score Model

**Appendix S1.** A brief introduction about sexual orientation and three domains: : sexual attraction, behavior, and identity

**1.1 Sexual attraction**

Sexual attraction refers to an individual's emotional, romantic, or sexual feelings towards others. It encompasses whom one is drawn to or desires, whether these feelings are fleeting or persistent. For some, attraction might be exclusively towards the opposite gender (heterosexual), the same gender (homosexual), or a combination of genders (bisexual). However, it's crucial to note that attraction is a spectrum, and individuals might experience degrees of attraction that don't fit neatly into these categories.

**1.2 Sexual behavior**

Sexual behavior concerns the physical acts individuals engage in with their partners, such as kissing, touching, or sexual intercourse. It's possible for one's sexual behavior to differ from their attractions or identity. For example, a person might identify as homosexual but engage in sexual activities with partners of the opposite gender due to societal pressures, personal circumstances, or other reasons. Analyzing behavior provides a tangible measure of sexual activity, but it doesn't always align seamlessly with the other dimensions of sexual orientation.

**1.3 Sexual identity**

Sexual identity is about how individuals perceive themselves and how they choose to label or define their sexual orientation. Common labels include "gay," "lesbian," "bisexual," "straight," and many others, but the list is ever evolving as society grows more understanding of the myriad ways people identify. This dimension is deeply personal, rooted in an individual's internal sense of self, and influenced by societal, cultural, and personal experiences. For some, their sexual identity aligns with their attraction and behavior, but for others, there might be discrepancies.

**Appendix S2.** Proof

**1. Without controlling for mediators**

The regression model

$$E[Y|q, x] = \beta_0 + \beta_1 * q + \beta_2 * x$$

**Notations**

Q: Sexual and Gender Minority (SGM) status; Q=1=SGM population; Q=0=heterosexual population
Y: Health outcome
X: Baseline social support level
$G(0)$: A random draw of heterosexual population's baseline social support level
$Y_x$: Counterfactual outcome when the baseline social support level is x
$Y_a$: Counterfactual outcome when the joint effect of SGM status is a
$E[Y_{G(0)}|Q = 1]$: The expected outcome in the SGM population, if we set their baseline social support level the same as a random draw of white population's baseline social support level

**Assumptions**

$E[Y|Q = 1, x] = E[Y_x|Q = 1]$
(Consistency and exchangeability: the expected outcome of the SGM population with baseline social support level as x, is the same as the expected potential outcome when the baseline social support is set as x in the SGM population; exchangeability is assumed by conditional on Q=1)

$E[Y_x|Q = 1] = E[Y_x|Q = 1, G(0) = x]$
(The expected counterfactual outcome of setting baseline social support level to x in the SGM population is the same, irrespective of conditional on a random draw of baseline social support level from a heterosexual population)

$\Pr(x|Q = 0) = \Pr(G(0) = x)$
(By definition of the notation)

$\Pr(G(0) = x) = \Pr(G(0) = x|Q = 1)$
($G(0)$ and $Q$ are independent)

**Causal interpretation from observed data**

$$\beta_1 = E[Y|Q = 1, x] - E[Y|Q = 0, x]$$

$$= \sum_x E[Y|Q = 1, x] * \Pr(x|Q = 0) - E[Y|Q = 0, x] * \Pr(x|Q = 0)$$

(Using $\Pr(x|Q = 0)$ as the weighting factor. We assume that the SGM population has the same baseline social support level distribution as the heterosexual population.)

$$= \sum_x E[Y|Q = 1, x] * \Pr(x|Q = 0) - E[Y|Q = 0]$$

$$= \sum_x E[Y_x|Q = 1] * \Pr(G(0) = x) - E[Y|R = 0]$$

$$= \sum_x E[Y_x|Q = 1, G(0) = x] * \Pr(G(0) = x|Q = 1) - E[Y|R = 0]$$

$$= E[Y_{G(0)}|Q = 1] - E[Y|Q = 0]$$

**Interpretation**

The SGM inequality would remain if the SGM population's baseline social support level were set equal to that of the heterosexual population.

**2. Controlling for mediators**

**2.1 Direct Effect**

**Notations**

Q: Sexual and Gender Minority (SGM) status; Q=1=SGM population; Q=0=heterosexual population
M: Subsequent social support level
Y: Health outcome
X: Baseline social support level
$Y_m$: Counterfactual outcome when subsequent social support level is m

$H_x(0)$: A random draw from heterosexual population's subsequent social support level when their baseline social support level is x

$Y_{H_x(0)}$: Counterfactual outcome when the subsequent social support level is set to the same value as a random draw from heterosexual population's subsequent social support level with their baseline social support level as x

$E[Y_{H_x(0)}|Q = 1, x]$: Expected health outcome for the SGM population with baseline social support level x, if their subsequent social support level is set to a value from the heterosexual population's subsequent social support level distribution corresponding to the same baseline social support level.

**Assumptions**

$E[Y|Q = 1, m, x] = E[Y_m|Q = 1, x]$
(The effects of M on Y are unconfounded conditional on Q and X)

$E[Y_m|Q = 1, x] = E[Y_m|Q = 1, H_x(0) = m, x]$
(The expected potential health outcome for an SGM individual should be the same as if their subsequent social support level were a random draw from the heterosexual population's distribution, given the same baseline social support level.)

$\Pr(m|Q = 0, x) = \Pr(H_x(0) = m)$
(By definition of the notation)

$\Pr(H_x(0) = m) = \Pr(H_x(0) = m|q, x) = \Pr(H_x(0) = m|Q = 1, x) = \Pr(H_x(0) = m|Q = 0, x)$
($H_x(0), q, x$ are independent)

**Causal interpretation from observed data**

$$\sum_m E[Y|Q = 1, m, x] * \Pr(m|Q = 0, x) - \sum_m E[Y|Q = 0, m, x] * \Pr(m|Q = 0, x)$$

(Using $\Pr(m|Q = 0, x)$ as the weighting factor. We assume the distribution of subsequent social support level (m) for the SGM and heterosexual individuals is the same, given they have the same baseline social support level (x).)

$$= \sum_m E[Y_m|Q = 1, H_x(0) = m, x] * \Pr(H_x(0) = m|Q = 1, x) - E[Y|Q = 0, x]$$

$$= E[Y_{H_x(0)}|Q = 1, H_x(0), x] - E[Y|Q = 0, x]$$

**Interpretation**

The racial inequality that would remain in a population with baseline social support level as x, for the SGM population in this total population, if we set the subsequent social support level of SGM population the same as the subsequent social support level distribution in heterosexual population.

## 2.2 Indirect Effect

**Notations**

Q: Sexual and Gender Minority (SGM) status; Q=1=SGM population; Q=0=heterosexual population
M: Subsequent social support level
Y: Health outcome
X: Baseline social support level
$Y_m$: Counterfactual outcome when subsequent social support level is m
$H_x(1)$: A random draw from SGM people's subsequent social support level when their baseline social support is x
$Y_{H_x(1)}$: Counterfactual outcome when the subsequent social support level is set to the same value as a random draw from SGM people's subsequent social support level with their baseline social support level as x
$E[Y_{H_x(1)}|Q=1,x]$: Expected outcome of SGM people with baseline social support level x when their subsequent social support level is set to the same value as a random draw from SGM people's subsequent social support level with their baseline social support level as x

**Assumptions**

$E[Y|Q=1,m,x] = E[Y_m|Q=1,x]$
(The effects of M on Y are unconfounded conditional on Q, X)

$E[Y_m|Q=1,x] = E[Y_m|Q=1, H_x(1)=m, x] = E[Y_m|Q=1, H_x(0)=m, x]$
$(Y_m, H_x(1), and\ H_x(0)\ are\ independent)$

$\Pr(m|Q=1,x) = \Pr(H_x(1)=m)$
(The definition of the notation)

$\Pr(H_x(1)=m) = \Pr(H_x(1)=m|q,x) = \Pr(H_x(1)=m|Q=1,x) = \Pr(H_x(1)=m|Q=0,x)$
($H_x(1)$, q, x are independent)

Similarly,

$$\Pr(m|Q = 0, x) = \Pr(H_x(0) = m) = \Pr(H_x(0) = m|q, x) = \Pr(H_x(0) = m|Q = 0, x) = \Pr(H_x(0) = m|Q = 1, x)$$

**Causal interpretation from observed data**

$$\sum_m E[Y|Q = 1, m, x] * \Pr(m|Q = 1, x) - E[Y|Q = 1, m, x] * \Pr(m|Q = 0, x)$$

(For the second part of the equation, we are giving the weight of m in heterosexual population with baseline social support level x to the SGM population)

$$= \sum_m E[Y_m|Q = 1, H_x(1) = m, x] * \Pr(H_x(1) = m|Q = 1, x) - \sum_m E[Y_m|Q = 1, H_x(0) = m, x] * \Pr(H_x(0) = m|Q = 1, x)$$

$$= E[Y_{H_x(1)}|Q = 1, x] - E[Y_{H_x(0)}|Q = 1, x]$$

**Interpretation**

The difference in health outcomes for the SGM population with a baseline social support level of x when comparing two scenarios: one where their subsequent social support level is set as that of the SGM population, and another where it is set as the subsequent social support level of the heterosexual population.

## 2.3 Total Effect

**Causal interpretation from observed data**

$$E[Y|Q = 1, x] - E[Y|Q = 0, x]$$

$$= \sum_m E[Y|Q = 1, m, x] * \Pr(m|Q = 1, x) - \sum_m E[Y|Q = 0, m, x] * \Pr(m|Q = 0, x)$$

$$= \sum_m E[Y|Q = 1, m, x] * \Pr(m|Q = 1, x) - \sum_m E[Y|Q = 1, m, x] * \Pr(m|Q = 0, x)$$

$$+ \sum_m E[Y|Q = 1, m, x] * \Pr(m|Q = 0, x) - \sum_m E[Y|Q = 0, m, x] * \Pr(m|Q = 0, x)$$

$$= \left\{ \sum_m E[Y|Q=1,m,x] * \Pr(m|Q=1,x) - \sum_m E[Y|Q=1,m,x] * \Pr(m|Q=0,x) \right\} +$$

$$\left\{ \sum_m E[Y|Q=1,m,x] * \Pr(m|Q=0,x) - \sum_m E[Y|Q=0,m,x] * \Pr(m|Q=0,x) \right\}$$

$$= Indirect\ Effect + Direct\ Effect$$

$$= \left\{ E\left[Y_{H_x(1)}|Q=1,x\right] - E\left[Y_{H_x(0)}|Q=1,x\right] \right\} +$$

$$\left\{ E\left[Y_{H_x(0)}|Q=1, H_x(0), x\right] - E[Y|Q=0,x] \right\}$$

**Table S1.** Comparison of Demographic and Clinical Characteristics Between Participants with and without Missing Values in the National Health Interview Survey (NHIS), 2020-2021

|  | Without missing values | | With missing values | | Comparison |
| --- | --- | --- | --- | --- | --- |
|  | SGM | Heterosexual | SGM | Heterosexual |  |
| **Age, mean (SD)** | 42.43 (17.18) | 53.69 (18.22) | 41.49 (16.27) | 53.40 (18.12) | $F = 0.006$, $P = 0.947$ |
| **Sex, n (%)** |  |  |  |  | $\chi^2 = 7.38$, $P = 0.06$ |
| Male | 854 (42.76) | 19821 (45.83) | 221 (46.23) | 5907 (45.97) |  |
| Female | 1141 (57.14) | 23424 (54.16) | 257 (53.77) | 6942 (54.03) |  |
| **Race, n (%)** |  |  |  |  | $\chi^2 = 111.17$, $P < 0.001$ |
| American Indian or Alaska Native | 61 (3.05) | 695 (1.61) | 8 (1.67) | 213 (1.66) |  |
| Asian | 74 (3.71) | 2516 (5.82) | 20 (4.18) | 725 (5.64) |  |
| Black or African American | 191 (9.56) | 4600 (10.64) | 43 (9.00) | 1356 (10.55) |  |
| Other | 67 (3.36) | 553 (1.28) | 11 (2.30) | 153 (1.19) |  |
| White | 1516 (75.91) | 32871 (76.00) | 375 (78.45) | 9733 (75.75) |  |
| **Educational Attainment, n (%)** |  |  |  |  | $\chi^2 = 111.17$, $P < 0.001$ |
| High School Graduate or Below | 426 (21.33) | 13977 (32.32) | 108 (22.64) | 4173 (32.62) |  |
| Bachelor's Degree | 1188 (59.49) | 22420 (51.84) | 265 (55.56) | 6631 (51.83) |  |
| Master's degree and above | 377 (18.88) | 6668 (15.42) | 100 (20.96) | 1933 (15.11) |  |
| **Family poverty ratio, mean (SD)** | 4.24 (3.09) | 4.31 (2.99) | 4.28 (3.01) | 4.33 (2.97) | $F = 0.453$, $P = 0.57$ |
| **Frequency of Social Support, n (%)** |  |  |  |  | Fisher's exact test, $P < 0.001$ |
| Always | 785 (39.31) | 23226 (53.71) | 0 | 16 (47.06) |  |
| Usually | 682 (34.15) | 11021 (25.48) | 0 | 6 (17.65) |  |
| Sometimes | 340 (17.03) | 4995 (11.55) | 0 | 2 (5.88) |  |
| Rarely | 113 (5.66) | 1564 (3.62) | 0 | 4 (11.76) |  |
| Never | 66 (3.30) | 1953 (4.52) | 0 | 3 (8.82) |  |
| **Changes in social support levels, n (%)** |  |  |  |  | Fisher's exact test, $P < 0.001$ |
| More social support | 409 (20.48) | 5216 (12.06) | 0 | 4 (11.76) |  |
| Less social support | 370 (18.53) | 4768 (11.03) | 0 | 2 (5.88) |  |
| About the same | 1209 (60.54) | 32809 (75.87) | 0 | 26 (76.47) |  |
| **Depression, n (%)** |  |  |  |  | $\chi^2 = 1053.3$, $P < 0.001$ |
| Yes | 848 (42.46) | 7356 (17.01) | 197 (41.21) | 2056 (16.00) |  |
| No | 1146 (57.39) | 35826 (82.84) | 281 (58.79) | 10774 (83.85) |  |

**Table S2.** Comparison of Demographic and Clinical Characteristics Between Respondents and Non-responses in the National Health Interview Survey (NHIS), 2020-2021

|  | Respondents | | Non-responses | | Comparison |
|---|---|---|---|---|---|
|  | SGM | Heterosexual | SGM | Heterosexual |  |
| **Age, mean (SD)** | 42.62 (17.30) | 54.06 (18.17) | 39.40 (14.83) | 48.32 (18.06) | $F = 0.38, P = 0.6$ |
| **Sex, n (%)** |  |  |  |  | $\chi^2 = 0.40, P < 0.05$ |
| Male | 801 (42.58) | 18491 (45.71) | 53 (45.69) | 1330 (47.57) |  |
| Female | 1080 (57.42) | 21959 (54.29) | 61 (52.59) | 1465 (52.40) |  |
| **Race, n (%)** |  |  |  |  | $\chi^2 = 129.34, P < 0.001$ |
| American Indian or Alaska Native | 59 (3.14) | 676 (1.67) | 2 (1.72) | 19 (0.68) |  |
| Asian | 74 (3.93) | 2455 (6.07) | 0 (0.00) | 61 (2.18) |  |
| Black or African American | 188 (9.99) | 4480 (11.08) | 3 (2.59) | 120 (4.29) |  |
| Other | 67 (3.56) | 545 (1.35) | 0 (0.00) | 8 (0.29) |  |
| White | 1493 (79.37) | 32294 (79.84) | 23 (19.83) | 577 (20.64) |  |
| **Educational Attainment, n (%)** |  |  |  |  | $\chi^2 = 798.52, P < 0.001$ |
| High School Graduate or Below | 389 (20.68) | 12542 (31.00) | 37 (31.90) | 1435 (51.32) |  |
| Bachelor's Degree | 1125 (59.81) | 21410 (52.93) | 63 (54.31) | 1010 (36.12) |  |
| Master's degree and above | 367 (19.51) | 6498 (16.06) | 10 (8.62) | 170 (6.08) |  |
| **Family poverty ratio, mean (SD)** | 4.29 (3.09) | 4.39 (3.00) | 3.39 (2.82) | 3.08 (2.51) | $F = 46.18, P = 0.021$ |
| **Frequency of Social Support, n (%)** |  |  |  |  | $\chi^2 = 454.96, P < 0.001$ |
| Always | 741 (39.39) | 22008 (54.41) | 44 (37.93) | 1218 (43.56) |  |
| Usually | 649 (34.50) | 10583 (26.16) | 33 (28.44) | 438 (15.67) |  |
| Sometimes | 323 (17.17) | 4680 (11.57) | 17 (14.66) | 315 (11.27) |  |
| Rarely | 109 (5.79) | 1469 (3.63) | 4 (3.45) | 95 (3.40) |  |
| Never | 59 (3.14) | 1710 (4.23) | 7 (6.03) | 243 (8.69) |  |
| **Changes in social support levels, n (%)** |  |  |  |  | $\chi^2 = 269.85, P < 0.001$ |
| More social support | 387 (20.57) | 4905 (12.13) | 22 (18.97) | 311 (11.12) |  |
| Less social support | 355 (18.87) | 4537 (11.22) | 15 (12.93) | 231 (8.26) |  |
| About the same | 1139 (60.55) | 31008 (76.66) | 70 (60.34) | 1801 (64.41) |  |
| **Depression, n (%)** |  |  |  |  | $\chi^2 = 867.29, P < 0.001$ |
| Yes | 808 (42.96) | 6997 (17.30) | 40 (34.48) | 359 (12.84) |  |
| No | 1073 (57.04) | 33453 (82.70) | 73 (62.93) | 2373 (84.87) |  |

## Table S3. STROBE Check List

STROBE Statement—checklist of items that should be included in reports of observational studies

|  | Item No. | Recommendation | Page No. | Relevant text from manuscript |
|---|---|---|---|---|
| Title and abstract | 1 | (*a*) Indicate the study's design with a commonly used term in the title or the abstract | 1 | Line 1-2 |
|  |  | (*b*) Provide in the abstract an informative and balanced summary of what was done and what was found | 1 | Line 4-18 |
| **Introduction** | | | | |
| Background/rationale | 2 | Explain the scientific background and rationale for the investigation being reported | 2 | Line 25-62 |
| Objectives | 3 | State specific objectives, including any prespecified hypotheses | 2 | Line 64-69 |
| **Methods** | | | | |
| Study design | 4 | Present key elements of study design early in the paper | 9 | Line 191-195 |
| Setting | 5 | Describe the setting, locations, and relevant dates, including periods of recruitment, exposure, follow-up, and data collection | 9-10 | Line 191-202 |
| Participants | 6 | (*a*) *Cohort study*—Give the eligibility criteria, and the sources and methods of selection of participants. Describe methods of follow-up<br>*Case-control study*—Give the eligibility criteria, and the sources and methods of case ascertainment and control selection. Give the rationale for the choice of cases and controls<br>*Cross-sectional study*—Give the eligibility criteria, and the sources and methods of selection of participants | 9-10 | Line 191-202 |
|  |  | (*b*) *Cohort study*—For matched studies, give matching criteria and number of exposed and unexposed<br>*Case-control study*—For matched studies, give matching criteria and the number of controls per case | NA | NA |
| Variables | 7 | Clearly define all outcomes, exposures, predictors, potential confounders, and effect modifiers. Give diagnostic criteria, if applicable | 9-11 | Line 204-241 |
| Data sources/ measurement | 8* | For each variable of interest, give sources of data and details of methods of assessment (measurement). Describe comparability of assessment methods if there is more than one group | 9-11 | Line 204-241 |
| Bias | 9 | Describe any efforts to address potential sources of bias | 11-12 | Line 243-261 |
| Study size | 10 | Explain how the study size was arrived at | NA | NA |
| Quantitative variables | 11 | Explain how quantitative variables were handled in the analyses. If applicable, describe which groupings were chosen and why | 10-11 | Line 204-241 |
| Statistical methods | 12 | (*a*) Describe all statistical methods, including those used to control for confounding | 11-12 | Line 243-261 |
| | | (*b*) Describe any methods used to examine subgroups and interactions | NA | NA |
| | | (*c*) Explain how missing data were addressed | 11 | Line 243-249 |
| | | (*d*) *Cohort study*—If applicable, explain how loss to follow-up was addressed *Case-control study*—If applicable, explain how matching of cases and controls was addressed *Cross-sectional study*—If applicable, describe analytical methods taking account of sampling strategy | 11-12 | Line 251-261 |
| | | (*e*) Describe any sensitivity analyses | 11-12 | Line 251-261 |
| **Results** | | | | |
| Participants | 13* | (a) Report numbers of individuals at each stage of study—eg numbers potentially eligible, examined for eligibility, confirmed eligible, included in the study, completing follow-up, and analysed | 12-13 | Line 265-273 |
| | | (b) Give reasons for non-participation at each stage | 12-13 | Line 265-273 |
| | | (c) Consider use of a flow diagram | 13 | Line 282 |
| Descriptive data | 14* | (a) Give characteristics of study participants (eg demographic, clinical, social) and information on exposures and potential confounders | 13 | Line 275-281 |
| | | (b) Indicate number of participants with missing data for each variable of interest | Supplementary material | Table S1, Table S2 |
| | | (c) *Cohort study*—Summarise follow-up time (eg, average and total amount) | NA | NA |
| Outcome data | 15* | *Cohort study*—Report numbers of outcome events or summary measures over time | NA | NA |
| | | *Case-control study*—Report numbers in each exposure category, or summary measures of exposure | NA | NA |
| | | *Cross-sectional study*—Report numbers of outcome events or summary measures | 13-14 | Line 291-316 |
| Main results | 16 | (*a*) Give unadjusted estimates and, if applicable, confounder-adjusted estimates and their precision (eg, 95% confidence interval). Make clear which confounders were adjusted for and why they were included | 13-14 | Line 291-316 |
| | | (*b*) Report category boundaries when continuous variables were categorized | NA | NA |
| | | (*c*) If relevant, consider translating estimates of relative risk into absolute risk for a meaningful time period | NA | NA |
| Other analyses | 17 | Report other analyses done—eg analyses of subgroups and interactions, and sensitivity analyses | 14-15 | Line 317-331 |
| Discussion | | | | |
| Key results | 18 | Summarise key results with reference to study objectives | 15-16 | Line 335-351 |
| Limitations | 19 | Discuss limitations of the study, taking into account sources of potential bias or imprecision. Discuss both direction and magnitude of any potential bias | 17 | Line 379-395 |
| Interpretation | 20 | Give a cautious overall interpretation of results considering objectives, limitations, multiplicity of analyses, results from similar studies, and other relevant evidence | 18-19 | Line 397-402 |
| Generalisability | 21 | Discuss the generalisability (external validity) of the study results | 18-19 | Line 397-402 |
| Other information | | | | |
| Funding | 22 | Give the source of funding and the role of the funders for the present study and, if applicable, for the original study on which the present article is based | NA | NA |

*Give information separately for cases and controls in case-control studies and, if applicable, for exposed and unexposed groups in cohort and cross-sectional studies.

Note: An Explanation and Elaboration article discusses each checklist item and gives methodological background and published examples of transparent reporting. The STROBE checklist is best used in conjunction with this article (freely available on the Web sites of PLoS Medicine at http://www.plosmedicine.org/, Annals of Internal Medicine at http://www.annals.org/, and Epidemiology at http://www.epidem.com/). Information on the STROBE Initiative is available at www.strobe-statement.org.

**Figure S1.** Comparison of Propensity Scores Between Sexual and Gender Minority (SGM) and Heterosexual (Straight) Participants in the National Health Interview Survey (NHIS), 2020-2021, With and Without the Inclusion of Subsequent Social Support in the Propensity Score Model

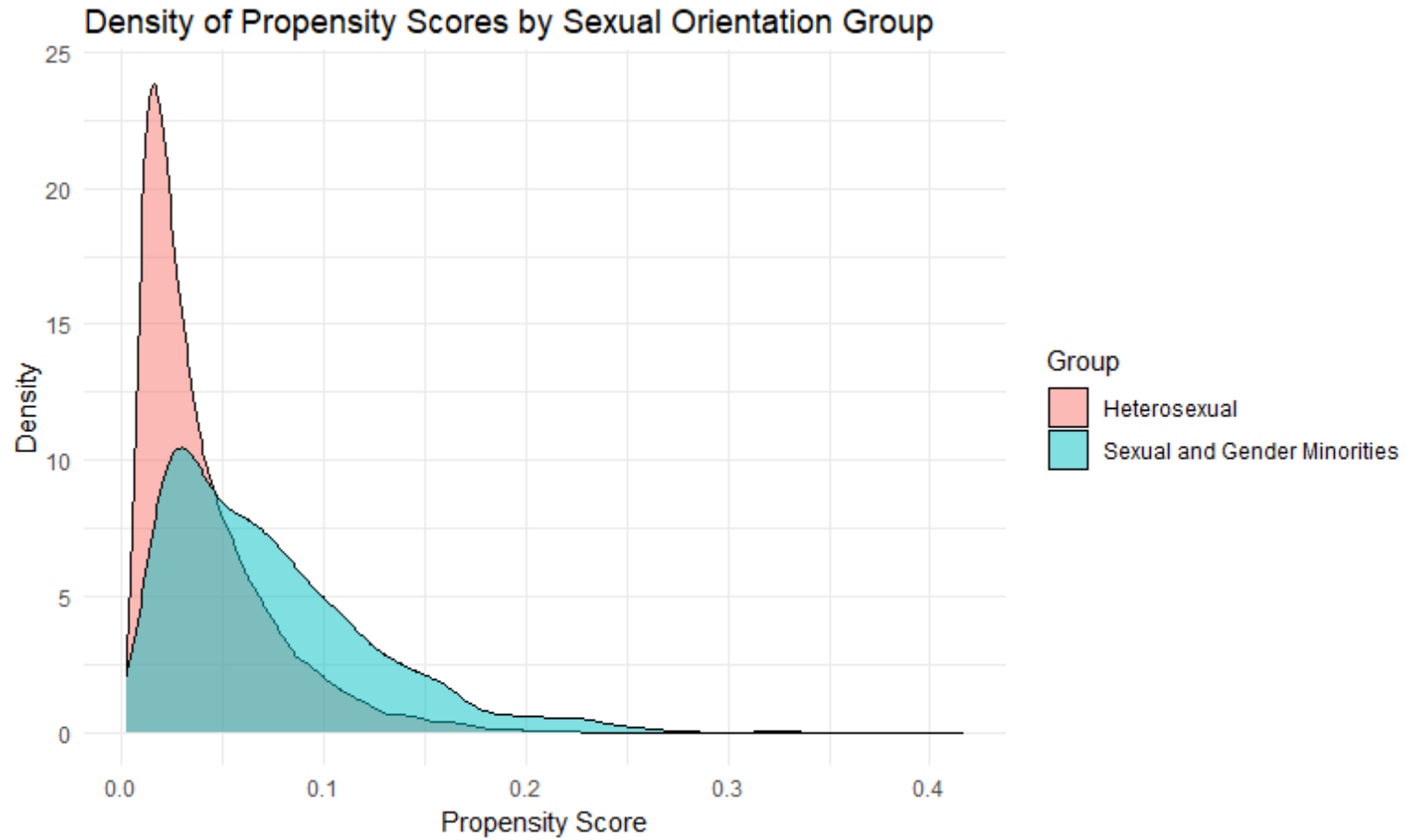

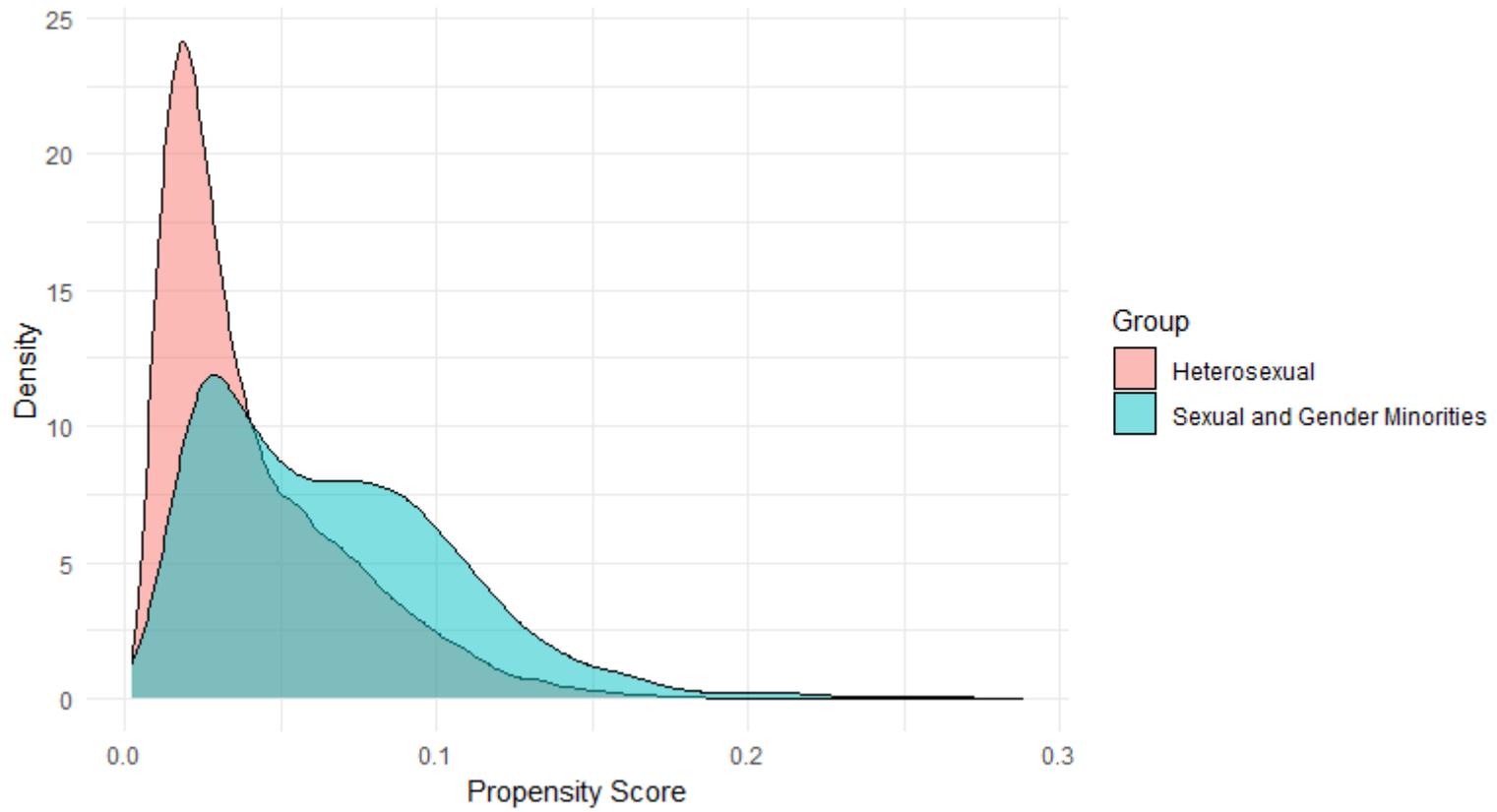